\documentclass[conference]{IEEEtran}
\IEEEoverridecommandlockouts
\usepackage{cite}
\usepackage{amsmath,amssymb,amsfonts}
\usepackage{algorithmic}
\usepackage{graphicx}
\usepackage{textcomp}
\usepackage{xcolor}
\usepackage{hyperref}
\usepackage{multirow}
\usepackage{booktabs}
\usepackage{threeparttable}

\usepackage{caption}
\usepackage{subfigure}

\def\BibTeX{{\rm B\kern-.05em{\sc i\kern-.025em b}\kern-.08em
    T\kern-.1667em\lower.7ex\hbox{E}\kern-.125emX}}
\begin{document}

\title{Contrastive Learning for Multi-Modal Automatic Code Review\\
\thanks{DOI reference number: 10.18293/SEKE2022-022} %Identify applicable funding agency here. If none, delete this.
}
%1\textsuperscript{st}
\author{\IEEEauthorblockN{Bingting Wu}
\IEEEauthorblockA{\textit{School of Computer Science and Technology} \\
\textit{Soochow University}\\
Suzhou, China \\
20204227028@stu.suda.edu.cn}
\and
\IEEEauthorblockN{Xiaofang Zhang}
\IEEEauthorblockA{\textit{School of Computer Science and Technology} \\
\textit{Soochow University}\\
Suzhou, China \\
xfzhang@suda.edu.cn}
}

\maketitle

\begin{abstract}
Automatic code review (ACR), aiming to relieve manual inspection costs, is an indispensable and essential task in software engineering. The existing works only use the source code fragments to predict the results, missing the exploitation of developer's comments. Thus, we present a Multi-Modal Apache Automatic Code Review dataset (MACR) for the Multi-Modal ACR task. The release of this dataset would push forward the research in this field. Based on it, we propose a Contrastive Learning based Multi-Modal Network (CLMN) to deal with the Multi-Modal ACR task. Concretely, our model consists of a code encoding module and a text encoding module. For each module, we use the dropout operation as minimal data augmentation. Then, the contrastive learning method is adopted to pre-train the module parameters. Finally, we combine the two encoders to fine-tune the CLMN to decide the results of Multi-Modal ACR. Experimental results on the MACR dataset illustrate that our proposed model outperforms the state-of-the-art methods.
\end{abstract}


\section{Introduction}
Code review is a critical part of software maintenance and evaluation. A general process of code review is shown in Fig. 1. Whenever the developer has submitted the revision code and the comments, the system will schedule an appropriate reviewer who will need to compare the differences between the original file, the modified file and combine the content of the developer's comments to verify whether the code meets the requirements. However, it also has a great demand for human resources \cite{sadowski2018modern}, which makes it impossible to expand code review on a large scale. 

Therefore, many researchers are committed to Automatic Code Review (ACR) to solve the problem \cite{shi2019automatic,siow2020core,Wu2022TureT}. For ACR, they provide the model with the original code and the revised code, and the model returns the suggestions on whether this modification is acceptable.

In this paper, we present a new perspective that the ACR should not only contain code fragments. The developer's comments, explaining why and how to change code, are also an essential source of information in the code review process. The model should fully consider the developer's comments when giving suggestions for code review. Therefore, we focus on the Multi-Modal ACR task. However, there are three main challenges: 1) there is no corresponding dataset, 2) most of the existing models ignore the problem of few-shot, and 3) there is no multi-modal model in the Multi-Modal ACR task.

\begin{figure}[t]
    \centering
    \includegraphics[width=1\linewidth]{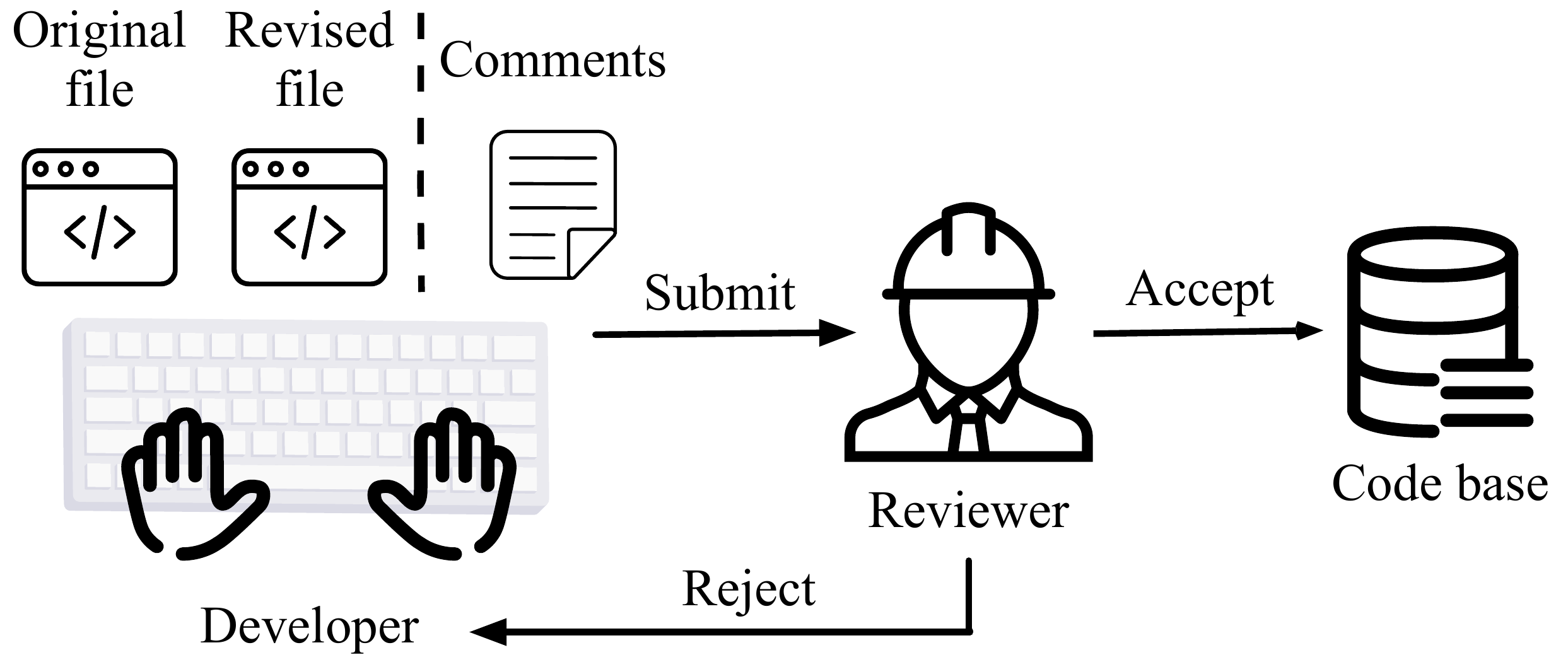}\\
    \caption{Traditional code review process.}
    \label{fig:cr}
\end{figure}

Since there is still a lack of relevant public datasets, we propose and open-source a Multi-Modal Apache Automatic Code Review dataset (MACR\footnote{https://github.com/SimAST-GCN/CLMN}), with 11 projects and over 500 thousand samples.

Recent works \cite{shi2019automatic,8812062,Wu2022TureT,siow2020core} have shown that deep learning methods perform better in capturing the syntactical and semantic information of the source code, enabling suitable code review suggestions. Among them, Shi et al. \cite{shi2019automatic} proposes a method called DACE, which uses long short-term memory (LSTM) and convolutional neural network (CNN) to capture the semantic and syntactical information, respectively. Zhang et al. \cite{8812062} proposes a method called ASTNN. The main idea is to obtain better feature extraction ability by dividing the abstract syntax tree (AST) into sentence-level blocks. Although these approaches are excellent in modeling code fragments, they do not consider the problem of few-shot. Misclassification may occur when the model encounters code fragments that have not been seen before. Therefore, in this paper, we adopt contrastive learning to obtain a uniform distributed vector representation to improve the robustness of the model.

%As for the absence of multi-modal methods, w
In addition, we explore a new method CLMN, to represent code snippets and textual content. We first model the code snippets and textual content separately using two independent encoding modules SimAST-GCN \cite{Wu2022TureT} and RoBERTa \cite{liu2019roberta}, respectively. Since RoBERTa already has ideal pre-training parameters, we need to pre-train SimAST-GCN. We adopt the method of contrastive learning to train the parameters of the SimAST-GCN module. After finishing the pre-training of the two encoders, we concatenate the vector representations of two encoders to obtain a joint representation of code and text to predict the final code review result.

Experiments results show that the proposed methods achieve significantly better results than the state-of-the-art baseline methods on MACR datasets. Our main contributions are summarized as follows: 
\begin{itemize}
  \item We provide a large-scale Multi-Modal Apache Automatic Code Review dataset (MACR) for the Multi-Modal ACR task.
  \item We propose a novel neural network CLMN to learn the combined relationship of the code fragments and textual content to improve the accuracy of Multi-Modal ACR.
  \item We conduct a series of large-scale comprehensive experiments to confirm the effectiveness of our approach. 
  \end{itemize}

The rest of this paper is organized as follows. Section II introduces the background. 
%ection III describes the construction process of the dataset. 
Section III describes our approach.
Section IV provides our experimental settings.
Section V presents the experimental results and answers the research questions.
%Section VI discusses several related works.
Finally, Section VI concludes our work.

\section{Background}
\subsection{Code Review}
The general process of traditional code review is shown in Fig. 1. In traditional approaches, researchers cannot solve the main challenge in code review: understanding the code \cite{bacchelli2013expectations}. Therefore, researchers can only improve efficiency from other aspects of code review, such as recommending suitable reviewers \cite{thongtanunam2015should,xia2017hybrid} and using static analysis tools \cite{balachandran2013reducing,diaz2013static}.

With the development of deep learning, we can understand the code by modeling the source code \cite{shi2019automatic,siow2020core,Wu2022TureT}. However, these methods only use code fragments to predict the results, lacking the exploitation of developer's comments in code review.

To solve the Multi-Modal ACR task, the model first extracts features from the original file, the revised file, and the comments. Then, the model needs to encode these features into vector representations and combine the code representation and the comment representation into a uniform representation. Finally, the neural network can make the final suggestion about the code review according to the uniform representation.

\subsection{Abstract Syntax Tree}
An abstract syntax tree (AST) is a kind of tree aimed at representing the abstract syntactic structure of source code. AST has been widely used as a vital feature in deep learning \cite{shi2019automatic,mou2014tbcnn,Wu2022TureT}. For example, it has a wide range of applications in source code representation \cite{8812062}, defect prediction \cite{shippey2019automatically}, and other fields. Each node of an AST corresponds to constructs or symbols of the source code. On the one hand, compared with plain source code, ASTs are abstract and do not include all details such as punctuation and delimiters. On the other hand, ASTs can be used to describe the lexical information and the syntactic structure of the source code. In this paper, we use the Simplified AST \cite{Wu2022TureT} as the input for the code encoder module.
%mou2014tbcnn removed at first cite in this subsection, due to paper limit
\iffalse
\subsection{SimAST-GCN}
In SimAST-GCN \cite{Wu2022TureT}, they parse the code fragment into the AST and use a specially designed algorithm to simplify the AST into a Simplified-AST. Then, they use preorder traversal algorithm to get the node sequence $\mathbf{x}=[\mathbf{x_1},\mathbf{x_2},...,\mathbf{x_n}]$ and the relation graph $\boldsymbol{A}\in \mathbb{R}^{n\times n}$, where $n$ is the number of nodes. Next, the representations for nodes are obtained by:
\begin{equation}
    \mathbf{H} = \{\mathbf{h}_1,\mathbf{h}_2,...,\mathbf{h}_n\} = \text{Bi-GRU}(\mathbf{x})
\end{equation}
Then, they use GCN to aggregate the information.% from their neighborhoods.
\begin{equation}
    \mathbf{h}^l = \mathrm{GCN}(\boldsymbol{A}, \mathbf{h}^{l-1})
\end{equation}
where $\mathbf{h}^l$ is the graph hidden status, $\mathbf{h}^0$ is the $\mathbf{H}$. Finally, they use a retrieval-based attention mechanism \cite{zhang2019aspect} to combine $\mathbf{H}$ and $\mathbf{h}^l$ to get the final representation of the code fragment.
\fi
\subsection{SimAST-GCN}
In our previous study, we proposed a method SimAST-GCN \cite{Wu2022TureT}. SimAST-GCN first parses the code fragment into the AST and uses a specially designed algorithm to simplify the AST into a Simplified-AST. Then, we use preorder traversal algorithm to get the node sequence $\mathbf{x}=[\mathbf{x_1},\mathbf{x_2},...,\mathbf{x_n}]$ and the relation graph $\boldsymbol{A}\in \mathbb{R}^{n\times n}$, where $n$ is the number of nodes. Next, the representations for nodes are obtained by:
\begin{equation}
    \mathbf{H} = \{\mathbf{h}_1,\mathbf{h}_2,...,\mathbf{h}_n\} = \text{Bi-GRU}(\mathbf{x})
\end{equation}
Then, we use GCN to aggregate the information from their neighborhoods.
\begin{equation}
    \mathbf{h}^l = \mathrm{GCN}(\boldsymbol{A}, \mathbf{h}^{l-1})
\end{equation}
where $\mathbf{h}^l$ is the graph hidden status, $\mathbf{h}^0$ is the $\mathbf{H}$. Finally, we adopt a retrieval-based attention mechanism \cite{zhang2019aspect} to combine $\mathbf{H}$ and $\mathbf{h}^l$ to get the final representation of the code fragment.

\subsection{Contrastive learning}
Contrastive learning aims to learn effective representation by pulling semantically close neighbors together and pushing apart non-neighbors. It assumes a set of paired samples $\mathcal{D}=\{(x_i,x_i^{+})\}_{i=1}^{m}$, where $x_i$ and $x_i^{+}$ are semantically related. We follow the contrastive framework in Chen et al. \cite{chen2020simple} and take a cross-entropy objective with in-batch negatives \cite{10.1145/3097983.3098202} : let $\mathbf{h}_i$ and $\mathbf{h}_i^{+}$ denote the representation of $x_i$ and $x_i^{+}$, the training objective for $(x_i,x_i^{+})$ with a mini-batch of $N$ pairs is:
\begin{equation}
    \ell = -\log{\frac{e^{sim(\mathbf{h}_i,\mathbf{h}_i^+)/\tau}}{\sum_{j=1}^N e^{sim(\mathbf{h}_i,\mathbf{h}_j^+)/\tau}}}
\end{equation}
where $\tau$ is a temperature hyperparameter and $sim(\mathbf{h}_1,\mathbf{h}_2)$ is the cosine similarity $\frac{\mathbf{h}_1^{\top} \mathbf{h}_2}{\lVert \mathbf{h}_1 \rVert \cdot \lVert \mathbf{h}_2 \rVert}$. In this work, we encode code fragments using a graph-based model SimAST-GCN and then train all the parameters using the contrastive learning objective (Eq. 3). For the text encoder, we use the pre-trained model RoBERTa and the parameters are trained by SimCSE \cite{gao2021simcse}. %: $\mathbf{h} = f_{\theta}(x)$

\iffalse
\subsection{Motivation}
In the field of ACR, previous work tends to use only the code fragments before and after modification to predict code review results. We believe that developers' comments are also crucial for the code review process. The comments explain the causes and effects of code modification. Therefore, it is necessary to incorporate the textual content into ACR.

In addition, in the ACR task, the model may encounter code fragments and annotation contents that have never been seen. Therefore, the spatial distribution of code representation and text representation should be as uniform as possible, so that the corresponding vector representation can be generated more accurately when encountering code fragments that have not been seen. Therefore, we adopt the method of contrastive learning to improve the overall robustness of the model.
\fi

\section{Proposed Approach}

\begin{figure*}[t]
    %\centering
    \begin{minipage}[t]{0.483\textwidth}
    \centering
    \raisebox{-0.05\height}{\includegraphics[width=1\linewidth]{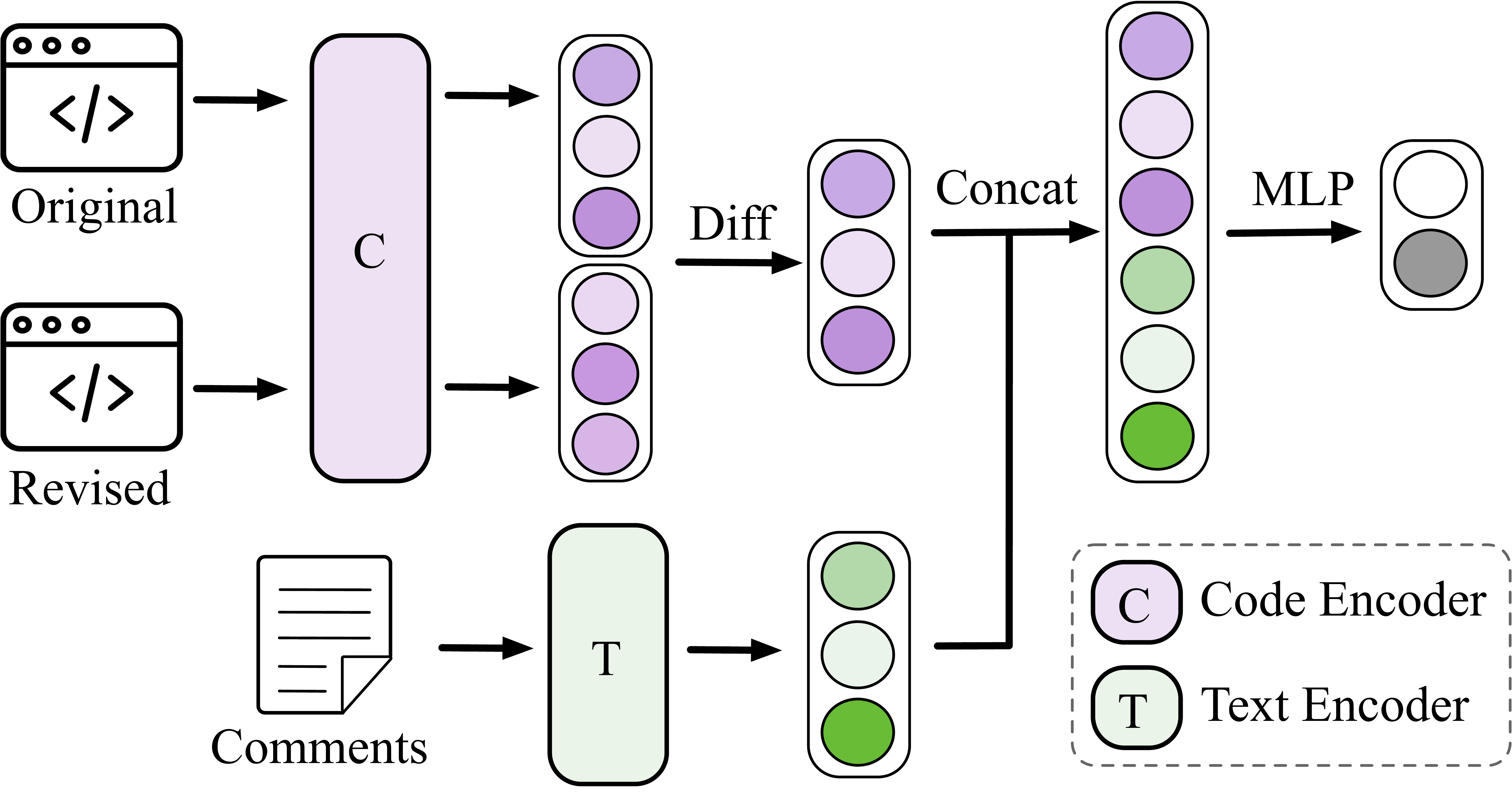}}
    \caption{General framework of CLMN.}
    \end{minipage}\hspace{9.pt}
    \begin{minipage}[t]{0.49\textwidth}
    \centering
    \raisebox{0.12\height}{\includegraphics[width=1\linewidth]{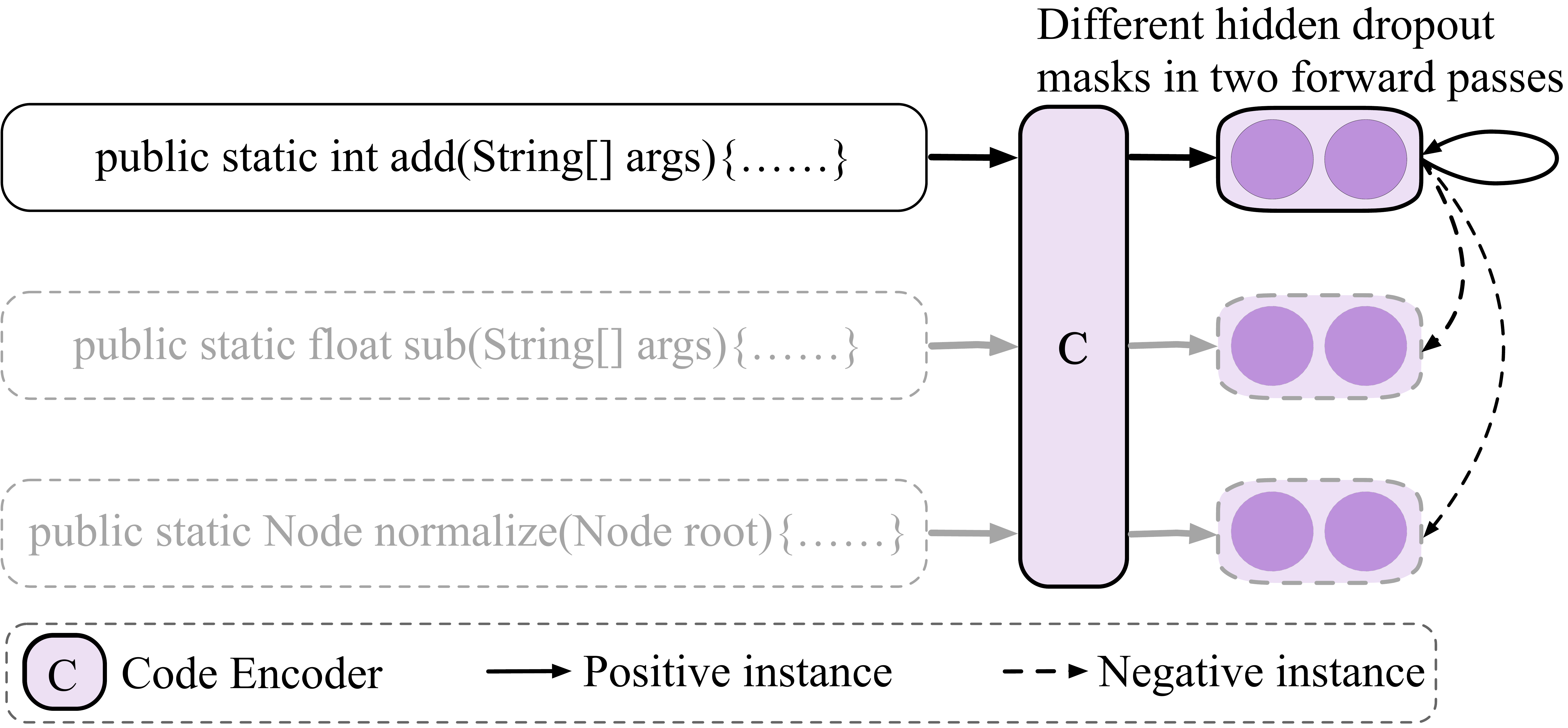}}
    \vspace*{-8mm}
    \caption{Unsupervised Code Encoder predicts the input code fragment itself from in-batch negatives, with different hidden dropout masks applied.}
    \end{minipage}
    %\includegraphics[width=0.48\linewidth]{clmn.pdf}
    %\caption{General framework of CLMN.}
    %\includegraphics[width=0.48\linewidth]{usp.pdf}
    %\caption{Unsupervised Code Encoder predicts the input code fragment \protect\\ itself from in-batch negatives, with different hidden dropout applied.}
    %\label{fig:clmn}
\end{figure*}

\iffalse
\begin{figure}[t]
    \centering
    \includegraphics[width=1\linewidth]{usp.pdf}\\
    \caption{Unsupervised Code Encoder predicts the input code fragment itself from in-batch negatives, with different hidden dropout applied.}
    \label{fig:usp}
\end{figure}
\fi

We introduce our approach CLMN in this section. As shown in Fig. 2, the framework of the proposed CLMN contains two main components: the Code Encoder (SimAST-GCN) and the Text Encoder (RoBERTa). For each encoder, we use contrastive learning to learn an adequate representation of code fragments and the developer's comments. For the Code Encoder, we use an unsupervised contrastive learning method to train its parameters. During the training process, we use dropout as noise to obtain positive samples. The process is shown in Fig. 3. For the Text Encoder, we use the parameters trained by SimCSE directly. After the unsupervised training for the two encoders, we use the pre-trained Encoders to fine-tune the finally model CLMN to predict the results.

\subsection{Contrastive Learning for Code Encoder}
The main challenge for contrastive learning is how to get positive samples. As shown in Fig. 3, in the Code Encoder, we pass the same code fragment to the model SimAST-GCN twice: by applying the standard dropout twice, we can obtain two different embeddings as ``positive pairs''. Then we take other code fragments in the same mini-batch as ``negative'' samples, and the model predicts the positive one among negative samples. Although it may appear strikingly simple, this approach is widely used in natural language models \cite{gao2021simcse}.

In this work, we take a collection of code fragments $\{x_i\}_{i=1}^m$ and use $x_i^+ = x_i$. We use the independent dropout for $x_i$ and $x_i^+$ to make it work. In SimAST-GCN, there are dropout placed on each GCN layers(default $p=0.1$). We denote $\mathbf{h}_i^z = f_\theta(x_i,z)$ where $z$ is a random mask for dropout. We feed the same input to the encoder twice and get two embeddings with different dropout masks $z,z'$, and the training objective of Code Encoder becomes: %\ell
\begin{equation}
    \mathcal{L} = -\log{\frac{e^{sim(\mathbf{h}_i^{z_i},\mathbf{h}_i^{z'_i})/\tau}}{\sum_{j=1}^N e^{sim(\mathbf{h}_i^{z_i},\mathbf{h}_j^{z'_i})/\tau}}}
\end{equation}
for a mini-batch of $N$ code fragments. Note that $z$ is just the standard dropout mask in SimAST-GCN and we do not add any additional dropout.

\subsection{CLMN}
\subsubsection{Encoder Module}
In our CLMN model, we have two pre-trained encoder modules, code encoder (SimAST-GCN) and text encoder (RoBERTa). For the code encoder, we denote $C = f(c)$ where $c$ is a code fragment and $C$ is the representation for this code fragment. For the text encoder, we denote $T = f(t)$ where $t$ is developer's comments and $T$ is the corresponding representation. In the Multi-Modal ACR task, each data example contains three parts: original code fragment, revised code fragment, and developer's comments. Thus, for each data, we can get three representations $\mathbf{C}^O,\mathbf{C}^R,\mathbf{T}^D$, corresponding to the original code fragment, revised code fragment, and the developer's comments.

\subsubsection{Combined Representation}
As shown in Fig. 2, after getting three representations, we first calculate the difference between the original code fragment and the revised code fragment:
\begin{equation}
    \mathbf{C} = \mathbf{C}^O - \mathbf{C}^R
\end{equation}
where $\mathbf{C}$ denotes the difference between the two code fragments. Then, we need to combine the difference representation with the comments representation, because the developer's comments describe the difference between the two code fragments.
\begin{equation}
    \mathbf{r} = [\mathbf{C},\mathbf{T}^D]
\end{equation}
Thus, we get the combined representation for this data sample.

\subsubsection{Prediction}
In this part, we make the prediction according to the combined representation $\mathbf{r}$.
\begin{equation}
    \mathbf{y} = \mathrm{softmax}(\mathbf{W} \mathbf{r} + \mathbf{b})
\end{equation}
where $\mathrm{softmax}(\cdot)$ is the softmax function to obtain the output distribution of the classifier, $\mathbf{W}$ and $\mathbf{b}$ are weights and biases, respectively.

%\subsection{Model Training}
The objective to train the classifiers is defined as minimizing the weighted cross entropy loss between the predicted and ground-truth distributions:
\begin{equation}
    \mathcal{L} = -\sum_{i=1}^S (w^O y_i log \hat{p_i} + w^R(1-y_i)log(1-\hat{p_i})) + \lambda\left\|\Theta\right\|_2
\end{equation}
where $S$ is the number of training samples, $w^O$ denotes the weight of incorrectly predicting a rejected change as approved, and $w^R$ denotes the weight of incorrectly predicting an approved change as 
rejected. These two terms provide the opportunity to handle an imbalanced label distribution. $\lambda$ is the weight of the $L_2$ regularization term. $\Theta$ denotes all trainable parameters.

\section{Experimental design}
This section introduces the process of the experiment, including the repository selection and the data construction, baseline setting, evaluation metrics, and experimental setting.
%to do here need to change a lot , and the paper should be at least 9 papers

\begin{table}[t]
\setlength{\belowcaptionskip}{-0.2cm} %段后
    \begin{center}
      \caption{Statistics of the MACR dataset.}
      \resizebox{1\columnwidth}{!}{
      \begin{tabular}{c|c|c|c}
        \hline
      
        \hline
        Repository & \#samples & \#rejected & reject rate \\
        \hline
        accumulo & 20,903 & 9,042 & 43.3\% \\
ambari & 41,424 & 14,262 & 34.4\% \\
beam & 82,438 & 22,141 & 26.9\% \\
cloudstack & 37,599 & 18,703 & 49.7\% \\
commons-lang & 9,958 & 8,959 & 90.0\% \\
flink & 133,614 & 112,733 & 84.4\% \\
hadoop & 39,958 & 33,861 & 84.7\% \\
incubator-pinot & 13,903 & 2,118 & 15.2\% \\
kafka & 64,872 & 37,077 & 57.2\% \\
lucene-solr & 58,823 & 50,042 & 85.1\% \\
shardingsphere & 19,993 & 1,216 & 6.1\% \\
        \hline
        
        \hline
        \end{tabular}
      }
    \end{center}
  \end{table}

\subsection{Dataset Construction}
We selected 11 projects from Github belonging to the Apache Foundation, which is a widely used code review source \cite{shi2019automatic,Wu2022TureT}. Three of them (\emph{accumulo, ambari, cloudstack}) were selected by Shi et al. \cite{shi2019automatic}. The remaining projects (\emph{beam, commons-lang, flink, hadoop, incubator-point, kafka, lucene-solr, shardingsphere}) were chosen because they have over 2000 stars. The language of all the projects is Java.

For data processing, we extracted all issues belonging to these projects from 2015 to 2020. Among these issues, many do not involve code submission, but only provide feedback, so we chose the types \emph{PullRequestEvent} and \emph{PullRequestReviewCommentEvent}.

In many practical cases, we can easily extract the original code, the revised code and the developer's comments from the issue, but because these codes are usually contained in many files, it is difficult for us to use them directly as the input of the network. Therefore, we assume that all of the changes are independent and identically distributed \cite{shi2019automatic}, so there is no connection between these changes, and if a file contains many changed methods, we can split these methods independently as inputs. 

Further, if we add a new method or delete a whole method, half of the input data is empty. So we discard these data because they cannot be fed into the network. That is, we only consider the case where the code has been changed. In addition, considering that the submitted data may be too large, we subdivide the code submitted each time into the method level for subsequent experiments. 

After processing the data, each piece of data comprises four parts: the original code fragment, the revised code fragment, the developer's comments and the label. The original code fragment and the revised code fragment are both method-level Java programs. The developer's comments are in English. The label uses 0 and 1 to represent rejection and acceptance. The basic statistics of the MACR are summarized in Table I.

In this paper, the rate of the rejection between 6\% and 90\% means that there is class imbalance during model training and it will lead to poor performance, so we use a weight-based class imbalance approach, setting the `class\_weight' parameter to `balance' in our model.

\subsection{Comparison Models}
In this paper, we compared our proposed model CLMN with three other models in the Multi-Modal ACR task. These models use different methods (including delimiter-based method, token-based method, tree-based method) to obtain the code features. The baseline models are as follows:

\begin{itemize}
\item \textbf{DACE} \cite{shi2019automatic} divides the code according to the delimiter and designs a pairwise autoencoder to compare the code. %recursive
\item \textbf{TBRNN} serializes the AST into tokens and uses an RNN to capture the syntactic and semantic information.
\item \textbf{ASTNN} \cite{8812062} splits large ASTs into a sequence of small statement trees and calculates the representation distance to compare the code.
\end{itemize}
Please note that for all the methods above, we add the same text encoder \textbf{RoBERTa} to make all the methods can get the developer's comments as the input.

In order to ensure the fairness of the experiment and the stability of the results, we ran all the methods on the new MACR dataset, and each experiment was repeated 30 times.

\subsection{Evaluation}
Since the Multi-Modal ACR can be formulated as a binary classification problem (accept or reject) \cite{shi2019automatic}, we choose the F1-measure (F1) because it is widely used and the Matthews correlation coefficient (MCC) because it can better evaluate the performance of a model on imbalanced datasets.
%we choose the commonly-used F1-measure (F1), Matthews correlation coefficient (MCC) as evaluation metrics.

The calculation formula of F1 is as follows:
\begin{equation}
	Precision=\frac{TP}{TP + FP}
\end{equation}
\begin{equation}
	Recall=\frac{TP}{TP + FN}
\end{equation}
\begin{equation}
	F1=2*\frac{Precision*Recall}{Precision+Recall}
\end{equation}
where TP, FP, FN, and TN represent True Positives, False Positives, False Negatives, and True Negatives, respectively. The F1 score is the harmonic mean of the precision and recall. The value range of F1 is [0,1].

The calculation formula of MCC is:
\begin{small} 
\begin{equation}
	MCC = \frac{TP \times TN - TP \times FN}{\sqrt{(TP+FP)(TP+FN)(TN+FP)(TN+FN)}}
\end{equation}
\end{small}
The value range of MCC is [-1,1]. For all metrics, higher values are better.

\subsection{Experimental Setting}
%All the experiments are conducted on a server with 24 cores of 3.8GHz CPU and a NVIDIA GeForce RTX 3090 GPU. 
In our experiments, we used the javalang tools\footnote{https://github.com/c2nes/javalang} to obtain ASTs for Java code, and we trained embeddings of symbols using word2vec with the Skip-gram 
algorithm. For the code encoder, the embedding size was set to 512. The hidden size of Bi-GRU was 512. The number of GCN layers was 4. For the text encoder, we used the default parameters. The coefficients $w^O$ and $w^R$ were related to the dataset, and the coefficient $\lambda$ of $L_2$ regularization item was set to $10^{-5}$. Adam was utilized as the optimizer with a learning rate of $10^{-5}$ to train the model, and the mini-batch was 64. We random initialized all the $\mathbf{W}$ and $\mathbf{b}$ with a uniform distribution.

All the experiments were conducted on a server with 24 cores of 3.8GHz CPU and a GeForce RTX 3090 GPU.

\begin{table}[t]
    \begin{center}
      \caption{Performance comparison in terms of F1.}
      \resizebox{1\columnwidth}{!}{
      \begin{tabular}{c|c c c|c}
        \hline
        
        \hline
        Repository & DACE & TBRNN & ASTNN & CLMN\\
        \hline
accumulo & 0.995 & 0.99 & 0.995 & \textbf{0.996}\\
ambari & 0.977 & 0.979 & 0.979 & \textbf{0.98}\\
beam & 0.995 & 0.994 & 0.995 & \textbf{0.996}\\
cloudstack & 0.979 & 0.979 & 0.98 & \textbf{0.981}\\
commons-lang & 0.957 & 0.963 & \textbf{0.965} & 0.964\\
flink & 0.421 & 0.959 & 0.972 & \textbf{0.979}\\
hadoop & 0.947 & 0.934 & 0.949 & \textbf{0.952}\\
incubator-pinot & 0.859 & 0.795 & 0.954 & \textbf{0.994}\\
kafka & 0.985 & 0.691 & 0.985 & \textbf{0.986}\\
lucene-solr & 0.978 & \textbf{0.979} & 0.978 & 0.975\\
shardingsphere & 0.992 & 0.991 & 0.927 & \textbf{0.994}\\
\hline
Average & 0.917 & 0.932 & 0.971 & \textbf{0.981}\\
        \hline

        \hline
        \end{tabular}
      }
    \end{center}
  \end{table}

\begin{table}[t]
    \begin{center}
      \caption{Performance comparison in terms of MCC.}
      \resizebox{1\columnwidth}{!}{
      \begin{tabular}{c|c c c|c}
        \hline
        
        \hline
        Repository & DACE & TBRNN & ASTNN & CLMN\\
        \hline
accumulo & 0.989 & 0.978 & 0.99 & \textbf{0.991}\\
ambari & 0.936 & 0.941 & 0.943 & \textbf{0.946}\\
beam & 0.982 & 0.977 & 0.983 & \textbf{0.985}\\
cloudstack & 0.957 & 0.957 & 0.96 & \textbf{0.961}\\
commons-lang & 0.954 & 0.96 & \textbf{0.962} & 0.961\\
flink & 0.306 & 0.952 & 0.967 & \textbf{0.975}\\
hadoop & 0.939 & 0.923 & 0.94 & \textbf{0.944}\\
incubator-pinot & 0.466 & 0.146 & 0.894 & \textbf{0.961}\\
kafka & 0.974 & 0.457 & 0.974 & \textbf{0.975}\\
lucene-solr & 0.975 & \textbf{0.976} & 0.975 & 0.971\\
shardingsphere & 0.876 & 0.863 & 0.295 & \textbf{0.899}\\
\hline
Average & 0.85 & 0.83 & 0.898 & \textbf{0.961}\\
        \hline

        \hline
        \end{tabular}
      }
    \end{center}
  \end{table}

\begin{table}[t]
    \begin{center}
      \caption{Impact of the addition of developer comments.}
      \resizebox{1\columnwidth}{!}{
      \begin{threeparttable}
      \begin{tabular}{c|c c|c c}
        \hline
        
        \hline
        \multirow{2}{*}{Repository} & \multicolumn{2}{c|}{CLMN w/o Text} & \multicolumn{2}{c}{CLMN} \\
        \cline{2-5}
         & F1 & MCC & F1 & MCC\\
        \hline
accumulo & 0.761 & 0.448 & \textbf{0.996} & \textbf{0.991}\\
ambari & 0.704 & 0.013 & \textbf{0.98} & \textbf{0.946}\\
beam & 0.824 & 0.321 & \textbf{0.996} & \textbf{0.985}\\
cloudstack & 0.532 & 0.215 & \textbf{0.981} & \textbf{0.961}\\
commons-lang & 0.469 & 0.474 & \textbf{0.964} & \textbf{0.961}\\
flink & 0.367 & 0.237 & \textbf{0.979} & \textbf{0.975}\\
hadoop & 0.393 & 0.298 & \textbf{0.952} & \textbf{0.944}\\
incubator-pinot & 0.673 & 0.181 & \textbf{0.994} & \textbf{0.961}\\
kafka & 0.604 & 0.298 & \textbf{0.986} & \textbf{0.975}\\
lucene-solr & 0.392 & 0.297 & \textbf{0.975} & \textbf{0.971}\\
shardingsphere & 0.839 & 0.191 & \textbf{0.994} & \textbf{0.899}\\
\hline
Average & 0.596 & 0.27 & \textbf{0.981} & \textbf{0.961}\\
        \hline

        \hline
        \end{tabular}
        \begin{tablenotes}
        %\footnotesize
        \scriptsize
        \item[*] CLMN w/o Text means the CLMN without the Text Encoder.
      \end{tablenotes}
        \end{threeparttable}
      }
    \end{center}
  \end{table}

\begin{table}[t]
    \begin{center}
      \caption{Impact of contrastive learning on model robustness.}
      \resizebox{1\columnwidth}{!}{
      \begin{threeparttable}
      \begin{tabular}{c|c c|c c}
        \hline
        
        \hline
        \multirow{2}{*}{Repository} & \multicolumn{2}{c|}{\emph{beam} based CLMN} & \multicolumn{2}{c}{\emph{self} based CLMN} \\
        \cline{2-5}
         & F1 & MCC & F1 & MCC\\
        \hline
accumulo & 0.996 & 0.99 & 0.996 & 0.991\\
ambari & 0.98 & 0.946 & 0.98 & 0.946\\
cloudstack & 0.98 & 0.961 & 0.981 & 0.961\\
commons-lang & 0.948 & 0.944 & 0.964 & 0.961\\
flink & 0.973 & 0.968 & 0.979 & 0.975\\
hadoop & 0.951 & 0.943 & 0.952 & 0.944\\
incubator-pinot & 0.989 & 0.934 & 0.994 & 0.961\\
kafka & 0.984 & 0.973 & 0.986 & 0.975\\
lucene-solr & 0.973 & 0.968 & 0.975 & 0.971\\
shardingsphere & 0.993 & 0.887 & 0.994 & 0.899\\
\hline
Average & 0.977 & 0.951 & 0.98 & 0.958\\
        \hline

        \hline
        %\multicolumn{5}{c}{\small *All the results are based on the pre-trained \emph{beam} Code Encoder. } \\
        \end{tabular}
        \begin{tablenotes}
        \scriptsize
        \item[*] \emph{beam} based CLMN: using the Code Encoder pre-trained by \emph{beam}.
        \item[*] \emph{self} based CLMN: using the Code Encoder pre-trained by themselves.
        %The cross-project results are based on the pre-trained \emph{beam} Code Encoder. So repository column doesn't have the \emph{beam} project.
        %\item[2] The quick brown fox jumps over the lazy dog.
      \end{tablenotes}
      \end{threeparttable}
      }
    \end{center}
  \end{table}

\section{Experimental Results}
This section shows the performance of our proposed method CLMN with other baseline methods. Therefore, we put forward the following research questions: 
\begin{itemize}
	\item RQ1: Does our proposed model CLMN outperform other models for Multi-Modal ACR?
	\item RQ2: Does the addition of developer's comments improve the performance of ACR?
	\item RQ3: How about the impact of contrastive learning on model robustness?
\end{itemize}

\subsection{Does our proposed model CLMN outperform other models for Multi-Modal ACR?}
Tables II, III show the comparison results on the MACR dataset. The results show that the proposed CLMN consistently outperforms all comparison models. This verifies the effectiveness of our proposed method at Multi-Modal ACR. 

Compared with DACE, our model CLMN achieves the best performance in both the F1 and MCC metrics, with 6.4\% and 11.1\% improvement, respectively. Compared with TBRNN and ASTNN, CLMN improved by 4.9\% and 1\% respectively in F1. In the MCC metric, CLMN also gained 13.1\% and 6.3\% improvement.% absolute

One aspect of our improvement comes from using a better code encoder. Our code encoder uses the Simplified AST as input, which is better than the source code fragment and the original AST. Simplified AST has fewer tokens and tighter node connections. Besides, our code encoder uses the GCN layers to get more syntactic and semantic information than the Bi-GRU layers in other methods.

Another aspect of our improvement comes from the application of contrastive learning. We utilize contrastive learning to obtain a more evenly distributed representation of the source code, which allows our model to perform better when encountering few-shot and 0-shot samples. In addition, the pre-training parameters obtained by contrastive learning can be flexibly applied to other tasks (discussed in RQ3), thereby reducing the consumption of resources.

In summary, this is the first study to apply contrastive learning to improve the performance in the multi-modal automatic code review task. Our experiments show the effectiveness of exploiting the Simplified AST and the contrastive learning.

\subsection{Does the addition of developer's comments improve the performance of ACR?}
Table IV shows the comparison results between the CLMN with and without the developer's comments. The results show that CLMN without the developer's comments tends to drop sharply.

We can observe that the CLMN without the developer's comments has the worst result on \emph{ambari}, with a 93\% drop in MCC compared to the original result. On the \emph{flink} project, the F1 value of CLMN without Text has dropped 61\%. On average, CLMN without Text has an overall decrease of 69\% in MCC and 38\% in F1. This establishes that Multi-Modal ACR can make better decisions with the developer's comments rather than only using the code fragments.

\subsection{How about the impact of contrastive learning on model robustness?}
To investigate the robustness of parameters obtained by contrastive learning, we use \emph{beam} as the original pre-training project. After getting the pre-trained parameters of the code encoder of \emph{beam}, we apply it to the remaining projects with fine-tuning to obtain the performance in other projects. Table V shows the impact of contrastive learning on model robustness.

We can find that after using \emph{beam's} pre-trained parameters, the performance of our model in the other ten repositories does not drop significantly compared to the performance pre-trained by themselves. There is even no drop in some repositories. On average, there is only a drop of 0.3\% in F1 and 0.7\% in MCC. In other words, the hyperparameters obtained using contrastive learning can be easily extended to different projects and obtain similar results. When the model encounters an unseen code fragment (few-shot or 0-shot), it can also generate its vector representation accurately.

The results are soundproof that contrastive learning can effectively increase the robustness of the model. It also demonstrates that the parameters obtained by contrastive learning can help the code get a more uniform distribution in the vector space, resulting in better performance across different projects.

\section{Conclusion}
In this paper, we first present MACR, which is the first open-sourced Multi-Modal dataset for Multi-Modal ACR. Then, we propose a novel model called CLMN. CLMN has two encoders, SimAST-GCN as the code encoder and RoBERTa as the text encoder. The two encoders are separately pre-trained using contrastive learning to obtain a uniformly distributed representation, which improves the robustness of the model and solves the few-shot and 0-shot issues. Finally, we combine the two encoders to obtain a vector representation of the code review and predict the results. Experimental results on the MACR dataset show that our proposed model CLMN outperforms state-of-the-art methods. Our code and experimental data are publicly available at \href{https://github.com/SimAST-GCN/CLMN}{https://github.com/SimAST-GCN/CLMN}.

\section*{Acknowledgment}
This work was partially supported by the National Natural Science Foundation of China (61772263, 61872177, 61972289, 61832009), the Collaborative Innovation Center of Novel Software Technology and Industrialization, and the Priority Academic Program Development of Jiangsu Higher Education Institutions.

\bibliographystyle{IEEEtran}
\bibliography{ref} 

\end{document}